\documentclass[5p]{elsarticle}

\usepackage{lineno,hyperref}

\usepackage{amsmath} 	   
\usepackage{siunitx} 	   
\usepackage{amsfonts}   
\usepackage{subcaption} 	   
\modulolinenumbers[5]

\journal{arXiv}

\newcommand{\xnf}{\mathbf{x}_\text{NF}}
\newcommand{\x}[1]{\mathbf{x}_#1}

\begin{document}
\begin{frontmatter}


\title{Time Series Fault Classification for \\Wave Propagation Systems with Sparse Fault Data}


\author[First,Second]{Erik Jakobsson}
\ead{erik.jakobsson@epiroc.com}
\author[Second]{Erik Frisk} 
\author[Second]{Mattias Krysander}
\author[First]{Robert Pettersson}

\address[First]{Epiroc Rock Drills AB, \"{O}rebro, 702 25, Sweden}
\address[Second]{Link\"{o}ping University, 
   Link\"{o}ping, 581 83, Sweden}

\begin{abstract}
In this work Time Series Classification techniques are investigated, and especially their applicability in applications where there are significant differences between the individuals where data is collected, and the individuals where the classification is evaluated. Classification methods are applied to a fault classification case, where a key assumption is that data from a fault free reference case for each specific individual is available. For the investigated application, wave propagation cause almost chaotic changes of a measured pressure signal, and  physical modeling is difficult. Direct application of One-Nearest-Neighbor Dynamic Time Warping, a common technique for this kind of problem, and other machine learning techniques are shown to fail for this case and new methods to improve the situation are presented. By using relative features describing the difference from the reference case rather than the absolute time series, improvements are made compared to state-of-the-art time series classification algorithms.  
\end{abstract}


\end{frontmatter}

\section{Introduction and motivation}
Time Series Classification (TSC) is an interesting field for the application of machine learning algorithms, and covers a wide range of algorithms from distance based classifiers \citep{abanda2019review} and feature based approaches \citep{helwig2015identification}, to more recent deep learning approaches, \citep{fawaz2019deep}. Such methods have potential to be used for fault classification from measurement data in many applications, including the application studied here, hydraulics in the mining sector. Automation and remote operation are trends driven by the desire to make mines safer by moving people away from machines and hazardous working environments. A logical side effect from this is a reduced awareness for operators and service personnel regarding the machines operating condition. There is potential for Time Series Classification to fill this gap. 

Numerous new Time Series Classification methods are published every year, often with the target to show improvement on benchmark time series data sets such as the UCR archive, \citep{dau2019ucr}, and many methods are able to handle time series from a wide range of applications. The work in this article does not intend to improve on such already good results on available time series classification benchmarks \citep{bagnall2017great}. Neither does it claim to have tested all possible variations, to rule out if one of them generalize enough to solve the problem. This work rather intends to discuss some particularities of how to approach an application, where training data from different individuals is scarce and does not cover the variability expected upon deployment. A key prerequisite is the availability of a small amount of nominal measurements from each individual upon deployment, a reasonable assumption for a condition monitoring application. The contribution of this work is a method how to use this reference data to improve classification from measurement data with a large impact from wave propagation.

Some background of this type of application is helpful to follow the reasoning. Hydraulic rock drills as seen in Fig.~\ref{fig:Boomer_MD20} are difficult to instrument due to a harsh environment including moisture and vibrations. Any sensor must be robust and reliable, leading to high costs and minimizing the number of sensors is of importance. For this work, a single pressure sensor is used, located on the inlet pressure line where many effects of internal conditions are believed to manifest. A hydraulic rock drill operates at frequencies where oscillations from pressure propagation is a phenomenon that needs to be taken into account. Further properties of the application include:

\begin{figure}[t]
\centering
\includegraphics[width=\columnwidth]{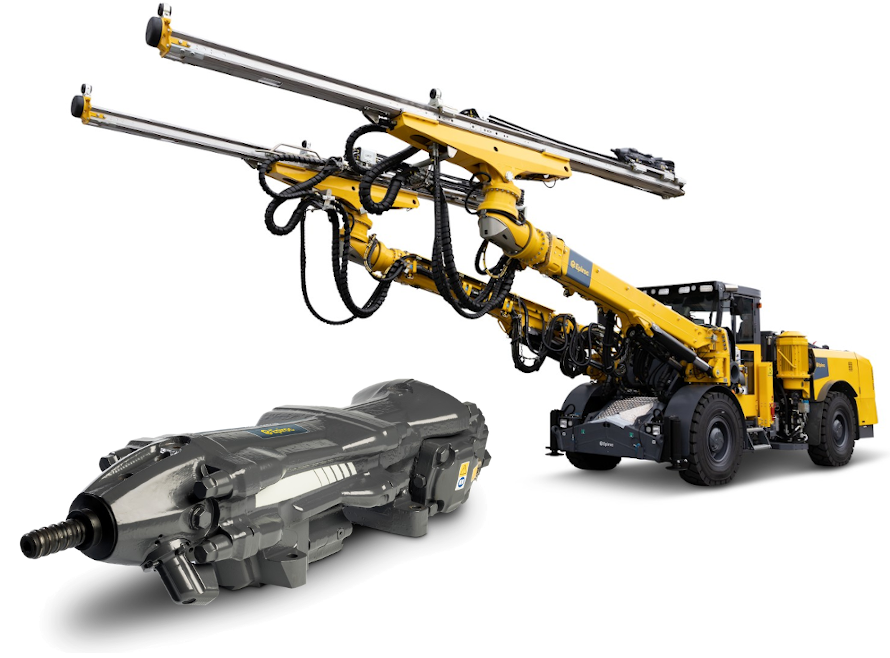}
\caption{A typical underground drill rig including the rock drill studied in this work. The rig carries two rock drills, positioned on the top right side of the feed beams. A close view of the rock drill is shown to the left.}  
\label{fig:Boomer_MD20}
\end{figure}

\begin{itemize}
\item[--]The measured signal is periodic, and governed by some cyclic phenomenon such as the repeated opening of a valve. 
\item[--]The fundamental machine frequency is influenced by various disturbancies, causing different events to occur at different times during a cycle depending on faults and individual variation such as unit configuration and manufacturing tolerances.
\item[--]Strong non-linearities, such as impacts and sudden valve openings.
\item[--]Wave propagation phenomenon occur, and since different events vary in time, it is difficult to predict how the superposition of the pressure waves occur.
\end{itemize}

The large influence from wave propagation on the pressure signatures has an interesting implication on the pressure measurement of the rock drill: It becomes sensitive to different boundary conditions, for example supply hose dimensions. From an implementation point of view, this is a problem since the same type of rock drill can be used on a variety of drill rigs, with varying feed systems, hose lengths, drilling directions, pump types etc. The number of possible combinations is large, and it is difficult to fully keep track of all possible variations. The sensitivity to boundary conditions means there will always be a significant difference between the systems used to obtain training data, and the system where the fault detection will be deployed. Variations between components further adds to this problem. 

Previous work \citep{jakobsson2021fault} showed how Time Series Classification, or more specifically Dynamic Time Warping \citep{sakoe1978dynamic} could be used to distinguish different faults by measuring percussion pressure, in a rock drilling application. The work showed promising results, but only looked at a single individual setup. It thus did not address the main problem discussed in this article, how the influence of individual variation severely reduce the performance of DTW and other Time Series Classification techniques and methods to cope with those variations. 

Different strategies can be used to address the problem. Either, one collects sufficient fault data from different faults in different individuals to find a model capable of generalizing sufficiently for unseen variations. This is the common machine learning approach. An alternative path is to build a physics based model based on first principles, and adapt it using data from specific individuals. For complex systems, such modeling can be both expensive and time consuming, if even possible. A third approach lies somewhere between the first two, to collect a limited set of data from just a few individuals, and then to adapt this model to specific individuals. Such ideas have been proposed with promising results in other domains, e.g., \citep{kiranyaz2015real}. Which approach that is more effective depends on the balance between complexity and what is easier to collect: Fault data from a representative population covering a sufficient part of the variability in data, or reference data from each specific individual setup upon deployment. In the fault detection case  the latter is arguably true, since each system can be considered fault free from start and thus the needed nominal calibration data can be collected from the deployed system. To induce faults in a large population is a huge undertaking for most applications.

This leads to the following research questions:
\begin{itemize}
\item[--]RQ1: How can classification be used when the target system is different from the training system?
\item[--]RQ2: Can No-fault data from the target system be used to reduce the effect of individual differences?
\item[--]RQ3: For this class of problems, how well do state-of-the-art Time Series Classification methods work?
\item[--]RQ4: What role does domain expertise have to make a machine learning scheme useful? 
\end{itemize}

The layout of the paper is as follows. First the application and its specific properties are presented. This is done to highlight how other applications with similar properties can benefit from the result. Then previous methodology, and its drawbacks are discussed, including the shortcomings of a model based approach to this problem. Finally, a new way to handle individual differences by utilizing reference No-fault data to incorporate information from the deployed system is presented.

\section{Detailed operation of the rock drill}
The rock drill is first described from a functional point of view, then with respect to the measured pressure oscillations during one impact cycle, i.e., the time from one impact to another.

\subsection{Functional description} 
\begin{figure}[t]
\centering
\includegraphics[width=\columnwidth]{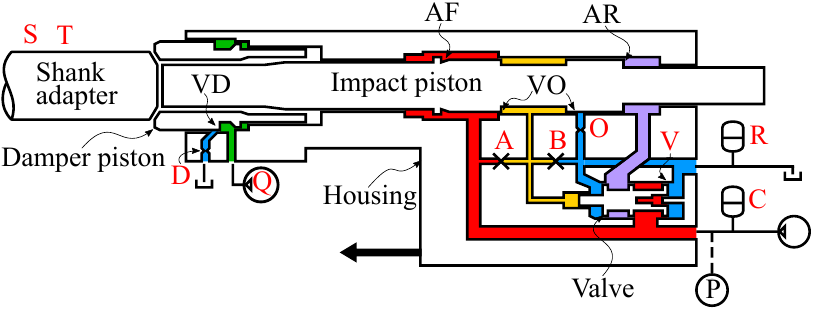}
\caption{A schematic view of the system studied in this work. Colored areas represent various hydraulic lines such as high pressure supply (red), low pressure return (blue), alternating pressure (purple), damper pressure (green) and control pressure for the valve (yellow). Different faults are shown as red letters.}  
\label{fig:md20}
\end{figure}

The rock drill is described with references to the schematic view of the rock drill as seen in Fig.~\ref{fig:md20}. The \textit{Impact piston} is forced to move in relation to the  \textit{Housing} by means of hydraulic pressure. The pressure is governed by a \textit{Valve}, whose position is in turn controlled by the \textit{Impact piston} location through two valve openings \textit{VO}. The timing of the valve openings is crucial for the rock drill performance.

At its leftmost position, the \textit{Impact piston} hits a \textit{Shank adapter}, protruding from the rock drill and via a drill rod connected to a drill bit. Through the repeated impacts by the \textit{Impact piston}, stress waves are generated crushing the rock at the drill bit. To achieve a good connection between the drill bit and the rock, a \textit{Damper piston} maintains a steady force on the \textit{Shank adapter}, and also dissipates some reflected energy originating from the bit-rock interaction.

The impact cycle starts after the \textit{Impact piston} just hit the \textit{Shank adapter}. The rearward acceleration phase starts, as the area \textit{AF} is pressurized and the area \textit{AR} is not. The \textit{Impact piston} moves rightwards until one of the valve openings \textit{VO} connect the control line (yellow) to return pressure (blue). The \textit{Valve} shifts position, pressurizing \textit{AR}. Since \textit{AR} $>$ \textit{AF}, the retardation phase starts, and continues until the \textit{Impact piston} reaches its rightmost position. At this point, the forward acceleration phase starts, where the \textit{Impact piston} is accelerated towards the \textit{Shank adapter}. Right before the impact, the valve openings \textit{VO} once again pressurize the control line (yellow), the \textit{Valve} shifts position, and the area \textit{AR} is depressurized to allow a new cycle to start after impact.

\begin{figure}[t]
\centering
\includegraphics[width=\columnwidth]{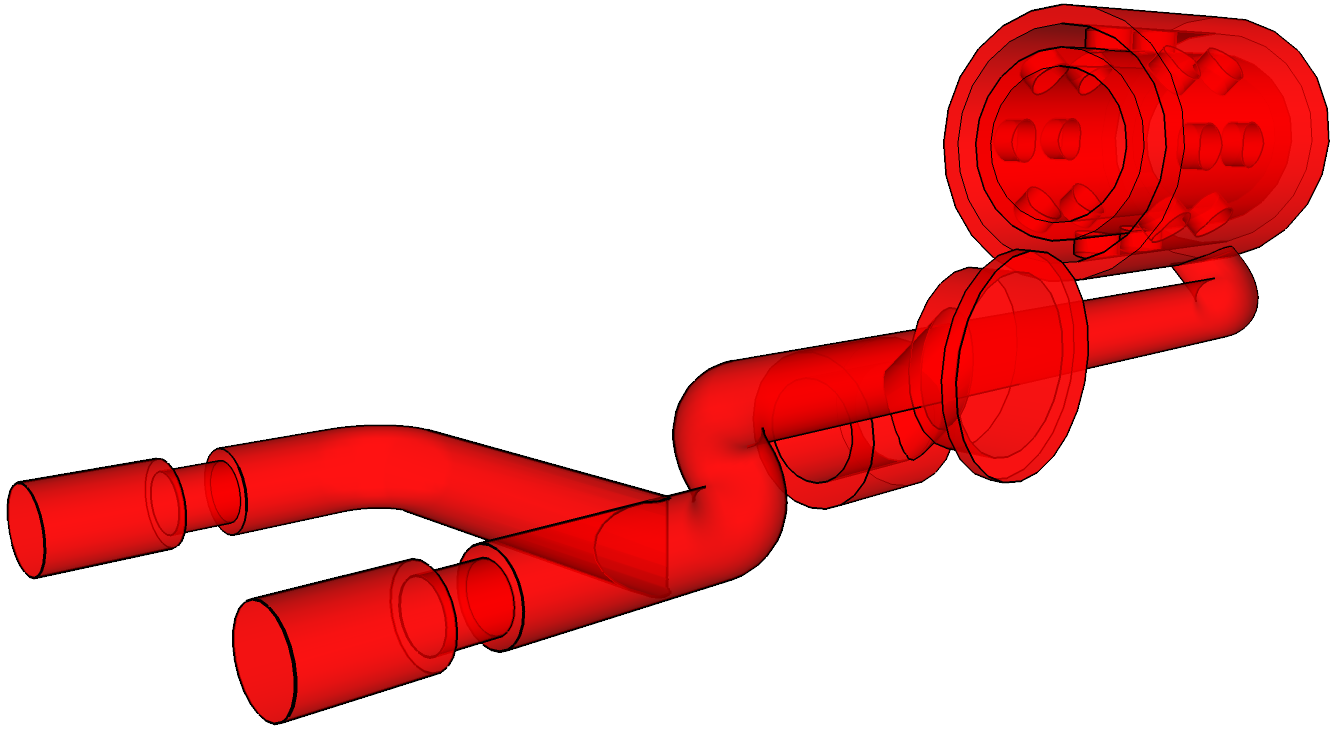}
\caption{The complexity of the high pressure inlet channel. To correctly simulate the percussion pressure, each bend, area change and volume need to be accounted for.}  
\label{fig:hp_inlet}
\end{figure}

The pressure sensor of interest in this work is located at the inlet port of the rock drill (P). This choice of sensor location is based on trying to obtain information from as many sources as possible within the rock drill, and the chosen position is hydraulically connected, at least momentarily, to all other high pressure locations in the percussion system of the rock drill. The measured pressure is largely governed by the different pressure wave propagation phenomena occurring in the rock drill. Any pressure change occurring in the system will propagate with the speed of sound in the hydraulic oil. A change of impedance, such as a change of cross sectional area of a line, will cause a part of the energy in the wave to be reflected, and some to be transmitted. Fig.~\ref{fig:hp_inlet} gives an idea of the complexity of the hydraulic channels. 

\subsection{Percussion pressure during one cycle}
The chaotic behavior of the measured pressure is a result of the interaction between wave propagation and the functionality of the rock drill. This section covers the different events that cause the pressure oscillations, to clarify the cause and to highlight the difficulties of modeling the system using first principles. 

\begin{figure*}
  \includegraphics[width=\textwidth]{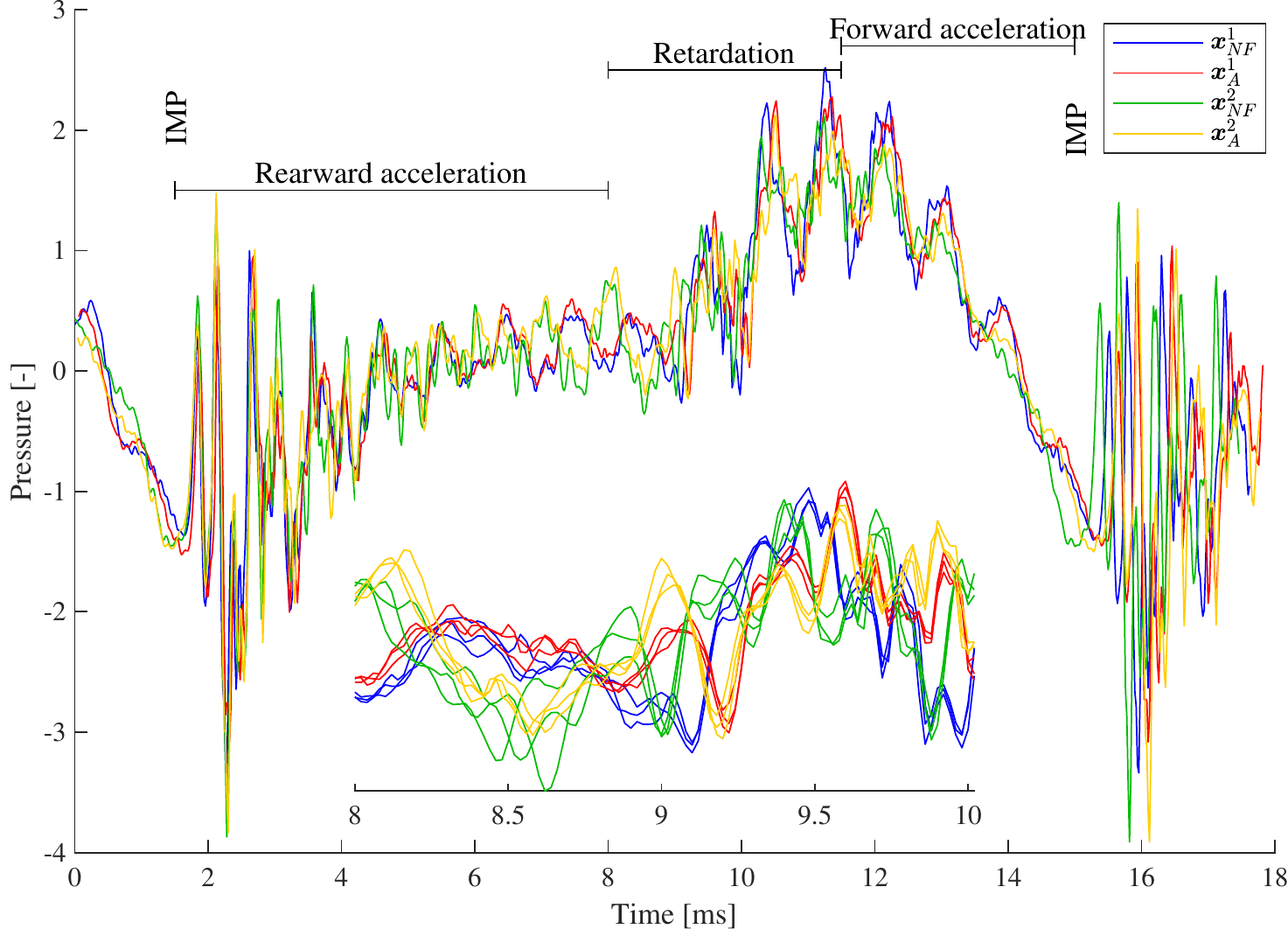}
  \caption{Normalized pressure during different phases of one impact cycle. The differences in pressure oscillations between a reference from individual one, $\xnf^1$, and fault $A$ from individual one, $\x{A}^1$, are more subtle than the reference from a different individual $\xnf^2$. The zoomed section shows how multiple strokes have very similar pressure signatures for the same individual, indicating that most oscillations are not random but rather a result of wave propagation. Around t = 9 ms, oscillations of fault $\x{A}^1$ starts to lag behind the $\xnf^1$ case. The same differences are observed for the second individual  $\x{A}^2$ compared to $\xnf^2$.}
\label{fig:timeseries}
\end{figure*}

Fig.~\ref{fig:timeseries} shows a number of detailed examples of the pressure measurements from different cycles. The pressure can be divided into a low-frequency trend, super positioned by high frequency oscillations. The low-frequency components of the pressure are caused by the supply and demand of hydraulic fluid. As the piston retracts after impact during the \textit{Rearward acceleration} phase, the supply system recovers to reach the nominal system pressure. When the main valve opens operation is switched to the \textit{Retardation} phase. Since the piston is moving rearward at this point, hydraulic oil is forced out of the rear chamber into the supply system, causing the pressure buildup seen. 

Eventually the piston comes to a halt and then starts moving forward. During this \textit{Forward acceleration} phase there is a gradual increase in demand of oil. As the piston accelerates, the supply system together with the accumulator supplies as much oil as is available, but the pressure still drops until the valve switches back and impact occurs, \textit{IMP}. The slower trend can be accurately modeled as seen in Fig.~\ref{fig:pinp_sim}.

The fast oscillations are caused by the sudden changes of flow in the system, triggering wave propagation in the channels and supply system. Four such events occur in the system. 

1. The closing of the valve, right before the impact occurs. The piston has reached its maximum velocity, and the percussion flow is high into the volume behind the piston. As the valve is closed, the inertia of the oil in the supply system causes a pressure peak. The pressure peak propagates through the hydraulic system, and cause a pressure oscillation governed mainly by the channel between the front most volume and the accumulator.

2. The opening of the valve, where oil is suddenly allowed to enter the currently low pressure volume behind the piston. This causes a sudden pressure drop locally, and the low pressure pulse propagates throughout the supply system.

3. The sudden stop of the piston upon impact, where the inertia of the oil flowing from the volume in front of the piston cause a pressure drop.

4. The opening of the control line, where oil is used to move the valve. This sudden flow causes a local pressure drop in the leftmost volume, which may or may not be large enough to be visible in the supply pressure at the sensor.

The fast changes are much more difficult to simulate accurately since the pulsations interact both with each other and with the changing geometry of the hydraulic system in the rock drill.

\section{Variations in the system}
The interaction from wave propagation in the system makes it sensitive to many types of changes. To highlight the benefit of using reference data from the same individual, different causes of difference in the pressure signals are categorized in the following way:

\textit{Noise} varies without control and can typically be reduced by averaging over longer time periods. The main source of noise is the response from the piston-shank interaction upon impact. Depending on the response from previous impacts, the piston can obtain different return velocities after impact. 
   
\textit{Control signals} are settings such as the set percussion pressure or feed pressure used to run the rock drill. It could also be the direction of drilling, or any other parameter that can be suddenly changed, but is known to the control system. 

\textit{Individual differences} are physical differences on the rock drill or the rig system that affect pressure signature. Individual differences include hose lengths, supply systems, drilling equipment, options on the rock drills, and manufacturing tolerances. 

\textit{Faults} are physical changes to the machine that are not intentional, not present for new machines, and not known.  Double faults are not considered in this work. Faults in rock drills can in turn be categorized in two types:
\begin{enumerate}
\item Faults that show no trace in the hydraulic circuit until complete failure.
\item Faults that affect performance.
\end{enumerate}

Item~1 includes faults like cracks growing in metallic parts. They are impossible to detect using the proposed method, since they leave no trace until complete failure of the part. 
For item~2 there is a wide range of faults. Some can be detected by a threshold on measured values, for example a completely empty high pressure accumulator, a blocked channel, or a broken pressure sensor. Other faults are more subtle but could cause secondary damage over time. A good example is a faulty low pressure accumulator that causes increased cavitation erosion damage. Excessive internal leakage could also have such effects and such faults are the target of this work.

\section{Problems with individual differences}
A previous work \citep{jakobsson2021fault} showed how One-Nearest Neighbor Dynamic Time Warping (1NN-DTW) could be successfully used to distinguish different fault modes from pressure data. The previous work did not take into account individual differences when evaluating detection performance of the approach. The following section highlights how 1NN-DTW and other methods here fail as such differences are introduced. 

Two individuals can have quite different pressure signatures due to a difference in supply hose length, as shown in Fig.~\ref{fig:timeseries}. The zoomed section at t~=~9~ms shows how an oscillation in time series $\x{A}^1$ starts to lag behind the reference. This is true also for the second individual, but would be very difficult to see without comparing to the second individuals reference case $\xnf^2$ . If only $\x{A}^1$ was part of the training data, it would thus be difficult to classify an unseen example from $\x{A}^2$. This poses a problem, since reference data for classification will originate from one or a few different individuals only. Most Time Series Classification algorithms rely on training data to cover all possible variations in data to generalize well and in many applications, such complete training data is difficult to come by.

\begin{figure}[t]
\centering
\includegraphics[width=\columnwidth]{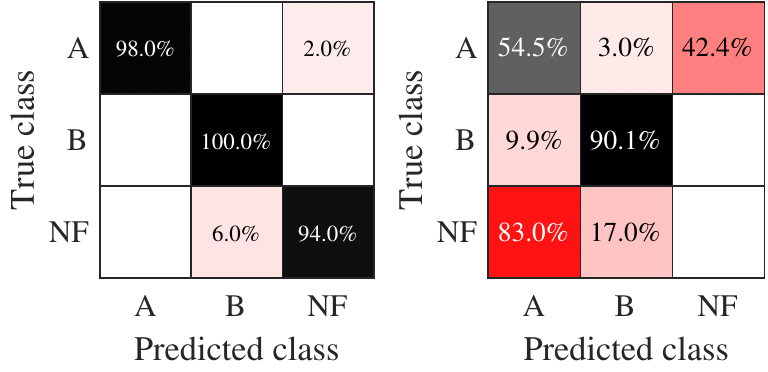}
\caption{1NN-DTW classification results. To the left, training and test data originate from the same individual. To the right, training and test data are from different individuals, where a hose length was changed between them. Both cases involve the same induced faults. The differences from a hose change cause the entire NF class to be wrongly classified as faulty.}  
\label{fig:failed_1NN_ABS}
\end{figure}

The effect from such a difference is shown in Fig.~\ref{fig:failed_1NN_ABS}. Training and testing a 1NN-DTW on the same individual results in almost perfect classification, but changing a hose length slightly between training and testing, results in entire classes being wrongly classified. A method to account for the individual differences is needed.    

\begin{figure}[t]
\centering
\includegraphics[width=\columnwidth]{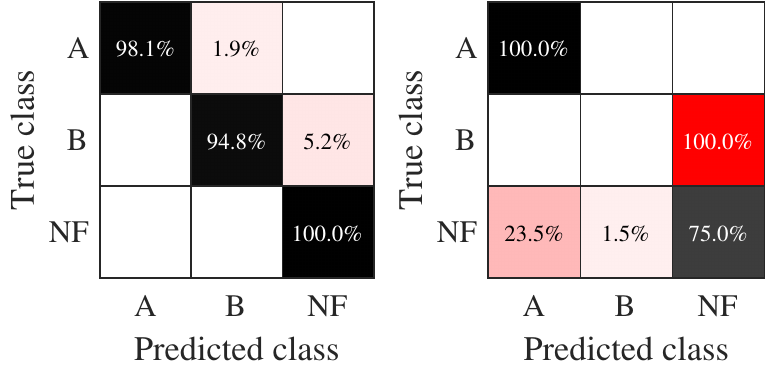}
\caption{Inception time fails in a similar way as 1NN-DTW when trained and evaluated on different individuals (right), with entire classes mis-classified compared to training and evaluating on the same class (left).}  
\label{fig:failed_inception_ABS}
\end{figure}

A number of other methods were also evaluated to gain confidence the problem was not with 1NN-DTW in particular, but rather with difference between individuals. From the vast number of time series classification methods, Inception Time \citep{fawaz2020inceptiontime} was chosen due to its recent success and relatively low training time. Fig.~\ref{fig:failed_inception_ABS} shows classification result for evaluation on both the same and different individuals. Just like 1NN-DTW, Inception Time mis-classifies entire classes though not necessarily the same ones. Similar results were obtained also for Shapelets \citep{ye2009time}, \citep{grabocka2014learning}, Rocket \citep{dempster2020rocket} and FLAG \citep{hou2016efficient}, strengthening the assumption that the problem lies not within the classifier, but rather in how to represent and address the individual differences.

\section{Simulation approach}
An attractive solution when data is scarce is a model based approach, based on a first principle physical model. Such methods reduce the need to collect data from multiple individuals and faults. Current state-of-the-art simulation models for rock drill simulations are based on Transmission Line Modeling \citep{krus1990distributed}, where the true three-dimensional geometry is represented by a large number of discrete components such as orifices, volumes, pipes, and hoses. Each component require approximations, such as where a line between volumes start and end, and how the pressure loss for a specific orifice behaves. 

\begin{figure}[t]
\centering
\includegraphics[width=\columnwidth]{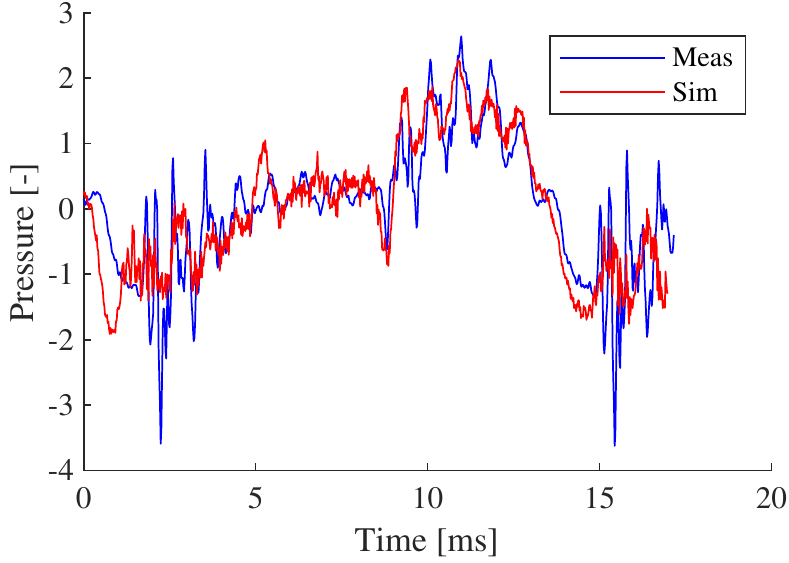}
\caption{Simulation accuracy captures the slow changes caused by supply/demand, but cannot accurately capture the various waves which superposition to the measured signals \textit{Meas}}  
\label{fig:pinp_sim}
\end{figure}

Fig.~\ref{fig:pinp_sim} shows results from the most accurate simulation model available to us. The slow changing pressures are simulated well, but the fast oscillations are not. This connects directly to the complexity of the input channel seen in Fig.~\ref{fig:hp_inlet}, and to the small differences that needs to be detected as shown in Fig.~\ref{fig:timeseries}. To accurately simulate the wave propagation, each and every change of impedance needs to be accounted for, and the pressure propagation need to be synchronized with the valve openings. 

Despite considerable effort and substantial domain knowledge including detailed geometry, and how to model the system, the model is not sufficiently accurate to detect faults where only small differences are observed in measurement data, for example the missing seal A in Fig.~ref{fig:timeseries}. Alternative ways to combine measurement data and domain knowledge are required, leading to data driven approaches. 

\section{Data}
Data is collected in a series of test-cell experiments with introduced faults and pre-defined individual variations. The measurement setup allows for some easily changed parameters such as pressures and flows. This makes it possible to do full factor trials for such parameters, and many such combinations are available in the data set. Other variations, such as induced faults or different hydraulic hoses require more manual labor. Hence less such combinations are available in the data set. 
Only one individual rock drill was available during trial, so differences from manufacturing tolerances could not be investigated, but are expected to be less prominent than differences from pressure and flow variations. To expand the useful data two controlled variables, percussion pressure and feed force, are chosen as alternatives to real individuals such as different rock drills or drill rig types. By doing so, a larger number of combinations between individuals and fault types can be investigated with the available data set. The changes in pressure signature and timing caused by these "pseudo-individuals" have a larger effect than real individual differences, and thus should be valid to show how the proposed methods are able to handle individual variation. The "pseudo-individuals" are called simply "individuals" from here on.

\begin{figure}[t]
\centering
\includegraphics[width=\columnwidth]{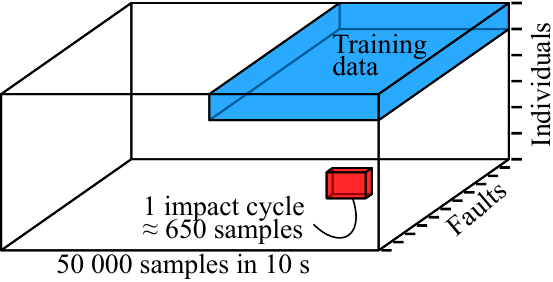}
\caption{The structure of the available data is shown. For each of the 11 faults, 6 different individuals are evaluated. From the 10s measurements, individual impact cycles are extracted for training and evaluation.}  
\label{fig:dataset}
\end{figure}

For a total of 11 induced fault, the combination of three percussion pressures and two feed forces give a total of six individual variations hence 66 cases. For each such case, ten seconds of data is collected, corresponding to approximately 800 impact cycles sampled at 50kHz. The ten second time series are divided into shorter segments, containing one impact cycle each. The following notation is used: Time series data sampled from impact cycle $k$ from individual $i$ and class $a$ will be denoted $\x{a}^{i,k}$. In cases where the impact cycle is not of interest, the corresponding index will be omitted. Fig.~\ref{fig:dataset} illustrates the available data.

From the six individuals, one is used for training and the remaining are used for test. The reason is to mimic a real scenario, were only a very low number of machines are used for collecting induced fault data and there is no large data set spanning all possible individual variations. Parts of the training data is held out to evaluate performance within the same individual. It is of importance to ensure that data from different individuals is not mixed, to prevent the classifier from bypassing the entire problem of individual differences. 

The different faults are selected based on possible occurrence and possibility to be detected using the pressure sensor. Included are also two changes to the drilling equipment, designated S and T in the list below. They are not really to be considered faults, but are included to show detection performance on other common changes in the drilling process. Each alteration is consistently referenced using a single capital letter, also visible in red in Fig.~\ref{fig:md20}. The following induced alterations are available in the data:
\begin{itemize}
\item[NF:]No-fault
\item[S:]Steel length, extra long.
\item[D:]Damper orifice is larger than usual.
\item[R:]Return accumulator, damaged.
\item[V:]Valve damage. A small wear-flat on one of the valve lands.
\item[Q:]Low flow to the damper circuit.
\item[C:]Charge level in high pressure accumulator is low.
\item[A:]A-seal missing. Leakage from high pressure channel to control channel.
\item[B:]B-seal missing. Leakage from control channel to return channel.
\item[T:]Thicker drill steel is used.
\item[O:]Orifice on control line outlet larger than usual.
\end{itemize}
The set of classes $\mathcal{F}$ is: $\mathcal{F}=\{\text{NF, S, D, R, V, Q, C, A, B, T, O}\}$.

\section{Method}
\begin{figure}[t]
\centering
\includegraphics[width=\columnwidth]{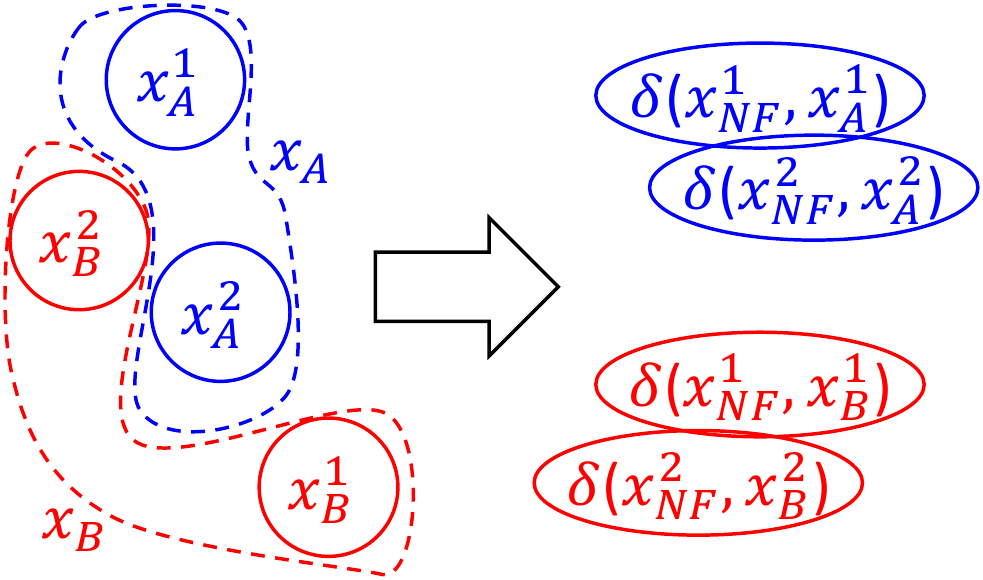}
\caption{Finding the true decision boundaries for $\x{A}$ and $\x{B}$ is hard with insufficient data, and using only individual~1 as training data results in misclassification of individual~2. By using reference data and transforming into feature space  $\delta$, classes become more invariant to individual differences and classification is possible.}  
\label{fig:invariant}
\end{figure}

As discussed, a common approach to handle individuals in machine learning is to train a model using data from a wide range of individuals and then to verify that the trained model becomes invariant to individual differences. Further, as discussed this approach can be difficult for a fault classification application where collecting induced fault data from a wide range of individuals is difficult. The following sections describe alternative methods as illustrated by Fig.~\ref{fig:invariant}, where a No-fault reference measurement from each individual is used to create a representation of the time series $\delta$ that is less affected by the individual differences.    

The representation is described as the $r$-dimensional output of a vector valued function $\delta$:
\begin{equation}
\delta:\mathbb{R}^{n_1} \times \mathbb{R}^{n_2} \rightarrow \mathbb{R}^r
\end{equation}
where $n_1$, $n_2$ are the lengths of the input time series and $r$ is the unknown length of the vector output, such that time series $\x{a}^i$ from the same class $a$ but different individuals $i,j$, in relation to reference data $\xnf$ give similar vector output:
\begin{equation}
\delta(\xnf^i,\x{a}^i) \approx \delta(\xnf^j,\x{a}^j) \quad \forall i,\forall j,\forall a, a \in F
\label{eq:rel_features_1}
\end{equation}
but different classes $a,b$ give different vector outputs
\begin{equation}
\delta(\xnf^i,\x{a}^i) \ne \delta(\xnf^i,\mathbf{x}_b^i) \quad \forall i, \forall a \ne b, a \in F, b \in F
\label{eq:rel_features_2}
\end{equation}
where lower case indices $a$ and $b$ represent any of the classes in set $F$. When calculating $\delta$ for the NF case, different impact cycles need to be used to avoid comparing two identical time series.

A straightforward representation would be the element-wise distance between a faulty case $\x{a}$ and the No-fault case $\xnf$.
\begin{equation}
\delta_{\text{euclidean}}(\xnf,\x{a})=|\xnf-\x{a}|
\end{equation}
To use such euclidean distance for wave propagation signals is neither suitable nor possible, since sections of the time signals are misaligned and of different length. Other ways to describe the difference are required, and the principle is first shown using a scalar valued hand crafted feature.

\subsection{Relative hand crafted features}
\label{sec:handcrafted}
\begin{figure}[t]
\centering
\begin{subfigure}[t]{\columnwidth}
\includegraphics[width=\columnwidth]{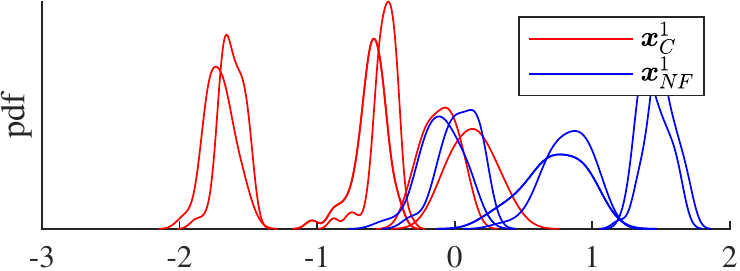}
\caption{The distribution of feature $P_{\text{drop}}$ is affected by individual differences, causing classes to be mixed up}
\label{fig:handcrafted_a}
\end{subfigure}
\begin{subfigure}[t]{\columnwidth}
\includegraphics[width=\columnwidth]{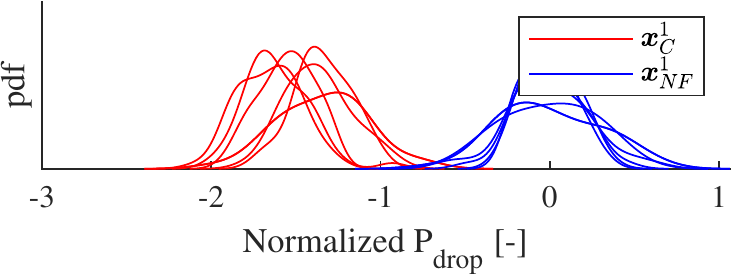}
\caption{When the bias is compensated for, classification becomes straightforward.}
\label{fig:handcrafted_b}
\end{subfigure}
\caption{Bias compensation for a scalar feature.}
\end{figure}

The previous work \citep{jakobsson2021fault} showed how hand crafted features were successful for some fault modes. The feature $P_{\text{drop}}$ is one such case, where the pressure change during the phase \textit{Forward acceleration} is measured, and a scalar is returned that has close correlation to the pre-charge pressure of the accumulator, fault C. Individual differences may affect also the performance of such approaches, for example by adding a bias to the features value. Fig.~\ref{fig:handcrafted_a} shows the distributions of this scalar feature for six individuals during No-fault $\xnf$ (blue) and low accumulator pressure $\x{C}$ (red). Individual differences adds bias to different individuals, causing classes to overlap. If only one of the NF distributions and C distributions were known, classification would not be possible for the overlapping cases. A key property is however that this bias on feature $P_{\text{drop}}$ is constant for a specific individual. By obtaining the feature's value for the No-fault case from each individual, the bias can be subtracted and classification becomes straightforward as seen in Fig.~\ref{fig:handcrafted_b}. This is analogous to creating a $\delta$ where the output vector has only one dimension according to:
\begin{equation}
\delta_{\text{pdrop}}:\mathbb{R}^{n_1} \times \mathbb{R}^{n_2} \rightarrow \mathbb{R}^1
\end{equation}  
where $\delta_{\text{pdrop}}$ is calculated as the difference of $P_{\text{drop}}$ between the faulty and No-fault case, for any combination of impact cycles: 
\begin{equation}
\delta_{\text{pdrop}} = P_{\text{drop}}(\x{C}^1) - P_{\text{drop}}(\xnf^1)
\end{equation}  
The result is shown in Fig.~\ref{fig:handcrafted_b}, where having any of the individuals with induced faults available for training would result in almost perfect classification for all the individuals through a threshold at $P_{\text{drop}}=-0.7$.

For a time series classification approach, this compensation is not as straightforward. It is unclear how to define $\delta$ for time series of difference lengths, and where different events occur at different locations. The following sections describe our approach on how to do this compensation when dealing with wave propagation data. But first, a short summary of Dynamic Time Warping. 

\subsection{Short summary: Dynamic Time Warping} 
Dynamic Time Warping \citep{sakoe1978dynamic} is a technique designed to measure similarity in time series with different speeds and different locations of characteristic sections, such as two different voices saying the same sentence. Two time series $\mathbf{x}(g), g \in \{1,\ldots,m\}$ and $\mathbf{y}(h), h \in \{1,\ldots,n\}$ are synchronized by choosing pairs of indices $(\mathbf{p}_\alpha,\mathbf{q}_\alpha)$ called a warping path $\mathbf{W}$:
\begin{equation}
{\mathbf{W}=(\mathbf{p}_1,\mathbf{q}_1),\ldots,(\mathbf{p}_\alpha,\mathbf{q}_\alpha),\ldots,(\mathbf{p}_k,\mathbf{q}_k)}
\label{eq:dtw_path}
\end{equation}
such that the DTW distance is minimized according to
\begin{equation}
DTW(\mathbf{x},\mathbf{y})= \min_{\mathbf{p}_\alpha, \mathbf{q}_\alpha} \sqrt{\sum^k_{\alpha=1}{(\mathbf{x}(\mathbf{p}_\alpha)-\mathbf{y}(\mathbf{q}_\alpha))^2}}
\label{eq:dtw}
\end{equation}
subject to the boundary conditions
\begin{equation} \label{eq:boundary_cond}
(\mathbf{p}_1,\mathbf{q}_1)=(1,1), (\mathbf{p}_k,\mathbf{q}_k)=(m,n)
\end{equation}
and a local step condition where for each position $(\alpha,\beta)$, the succeeding possible position is either of 
\begin{equation}
(\alpha+1,\beta),(\alpha,\beta+1),(\alpha+1,\beta+1).
\end{equation}

The output is a scalar measurement of distance between the two vectors, where some shifting in time is allowed to match up peaks and valleys of the signals. DTW is typically used as a distance measure in a one-nearest-neighbor approach, where samples are classified according to the most similar time series in a training data set. When comparing time series using such 1NN-DTW from different individuals, there is no simple notion of relative change as in the case of the hand-crafted feature in Section~\ref{sec:handcrafted}. A alternative is proposed in the following section. 

\subsection{Relative DTW measures}
\label{section:rel_DTW}
The objective is to describe relative changes from the No-fault case for a section of the measured pressure, similar to the bias subtraction performed for the hand crafted feature. A method to describe change from the reference signal while allowing some freedom to match oscillations is shown in this section.

The standard DTW formulation gives a scalar pseudo-metric for similarity of two time series, but describing the difference as only a distance to the reference is insufficient to distinguish different faults. A richer representation, such as a notion of where and how the time series differs from the No-fault case is needed. Two different methods are proposed to categorize the difference: amplitude difference $\delta_{\text{amp}}$ and time shift $\delta_{\text{ts}}$.

\begin{figure}[t]
\centering
\includegraphics[width=\columnwidth]{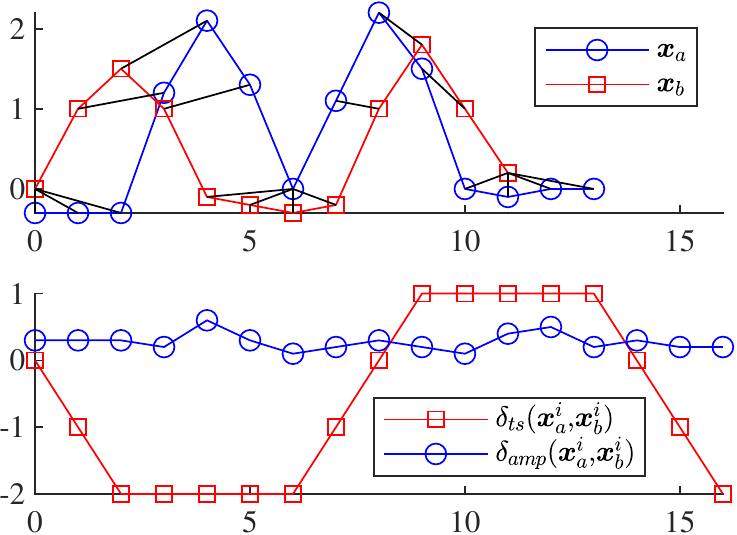}
\caption{Creation of the features measuring amplitude difference and time shift difference for the DTW aligned time series. The lower figure shows the feature values, calculates as the vertical and horizontal components of the matching-lines in black in the upper figure.}  
\label{fig:dtw_timeshift}
\end{figure}

For $\delta_{\text{amp}}$ the target is to capture amplitude differences that result from different damping in the hydraulic circuit, without a need to manually synchronize different segments of the time series. First the standard DTW algorithm is used to align the two vectors. Through the alignment, each point in respective time series is mapped to one or more points in the other time series as illustrated by the black lines in Fig.~\ref{fig:dtw_timeshift}. In this way the main oscillations in the pressure signals are lined up, and the amplitude differences measured are not caused by misalignment. Instead of using (\ref{eq:dtw}) to find a scalar distance, a vector difference is defined according to:

\begin{equation}
\delta_{\text{amp}}(\xnf,\x{a})= |(\xnf(\mathbf{p}_\alpha)-\x{a}(\mathbf{q}_\alpha))|
\end{equation}
 
where the vector $\delta_{\text{amp}}$ constitutes of a signature for how the curves differ, a vector version of the DTW measure of similarity and $\xnf(\mathbf{p}_\alpha)$ where $\x{a}(\mathbf{q}_i)$ are the warped time series. The length of $\delta_{\text{amp}}$ is the same as for the warping path and changes depending on how similar the curves are. 

For $\delta_{\text{ts}}$, the goal is to capture timing differences between the signals. Fig.~\ref{fig:dtw_timeshift} shows the conceptual creation of such a descriptive vector, where the feature values correspond to the horizontal component of each black line from the DTW matching. This corresponds to how many samples the time series $\x{b}$ needs to be shifted with respect to the reference time series $\x{a}$ to achieve optimal alignment. The difference is calculated as:

\begin{equation}
\delta_{\text{ts}}(\xnf,\x{a})= (\mathbf{p}_\alpha-\mathbf{q}_\alpha)
\end{equation}

where $\mathbf{p}_\alpha$ and $\mathbf{q}_\alpha$ are the warping path indices for the optimal alignment from (\ref{eq:dtw_path}).

\begin{figure}[t]
\centering
\includegraphics[width=\columnwidth]{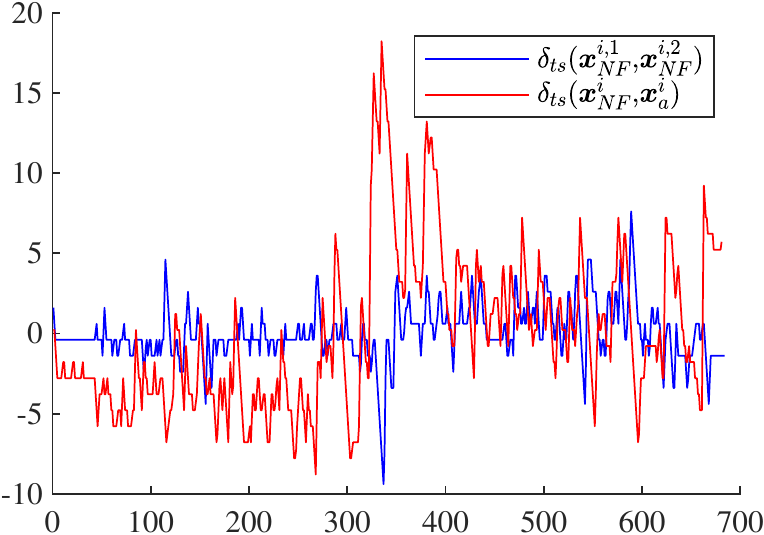}
\caption{Example of how the $\delta_{\text{ts}}$ feature captures differences in valve transition timing. A time shift difference is seen at index 330, required to match the faulty case to the No-fault reference. This causes $\delta_{\text{ts}}(\xnf^i,\x{a}^i)$ to look fundamentally different in shape from $\delta_{\text{ts}}(\xnf^{i,1},\xnf^{i,2})$, where the different indices $1$, $2$ indicate different cycles from the same individual $i$.}  
\label{fig:timeshift_A_NF}
\end{figure}

The resulting vector provides a signature for where and how much during a stroke the signal is shifted in time. For our kind of data, this feature can for example capture the difference between a late valve transition and the No-fault case, as seen in Fig.~\ref{fig:timeshift_A_NF}.

The vectors $\delta$ are similar to time series, and thus time series classification techniques are still used for training and evaluation. The most straightforward method would be Euclidean distance, but it is unsuitable due incapability to handle both slight shift i position of events, and also the varying length of the $\delta$-vectors. Several other methods are evaluated. First, 1NN-DTW, used with the different relative feature vectors as input. Second, a method proposed by \cite{kate2016using}, where multiple DTW distances to different training samples are used as features, described below. Third, the Inception Time network, using the various relative feature vectors. 

\subsection{Pairwise distances as features}
Pairwise distances \citep{kate2016using} is an extension of 1NN-DTW, where instead of finding only the closest training time series, a vector $\delta_\text{pd}$ containing distances to multiple different training time series is used for classification. Such vectors are used first to train a machine learning model from the training data, before deployment, and then to classify unknown cycles. The vector is created according to: 
\begin{equation}
\delta_\text{pd}=(p_{\text{amp}}^\text{NF},p_{\text{amp}}^\text{A},p_{\text{amp}}^\text{B},\ldots,p_{\text{ts}}^\text{NF},p_{\text{ts}}^\text{A},p_{\text{ts}}^\text{B},\ldots)
\label{eq:delta_pd}
\end{equation}
where each element represents a distance between the cycle to be classified to one class in the training data set. The method allows for use with non-distance types of features, for example the rock drill frequency, and provides a convenient way to use several of the proposed distance measures, such as $\delta_{\text{amp}}$ and $\delta_{\text{ts}}$, simultaneously. This is a main benefit for this technique, and is the reason it is evaluated in this work.

The process to generate the concatenated vector for a single cycle is described below, and is used both for training and evaluation data.
\begin{enumerate}
\item Calculate feature vectors $\delta_{\text{amp}}$ and $\delta_{\text{ts}}$ according to Section~\ref{section:rel_DTW} for all the single cycle and for all different classes $F$ from the training individual.
\item Choose \textit{n} vectors $\delta_{\text{amp}}$ from each class from the training individual to be references, denoted \textit{ref}.
\item Calculate DTW distance between the single $\delta_{\text{amp}}$ to each of the chosen \textit{n} reference vectors $\delta_{\text{amp}}$.
\item Average over all $n$ references within the same class. 
\item Repeat 2 to 4 for vectors $\delta_{\text{ts}}$ and any other feature to be used.
\item Concatenate the averages according to~(\ref{eq:delta_pd}).
\end{enumerate}

An example for a single position in vector $\delta_\text{pd}$ is shown below:
\begin{equation}
p_{\text{amp}}^A=\frac{1}{n} \sqrt{\sum^n_{s=1} \text{DTW}(\delta_{\text{amp}}(\xnf^\text{ref,s},\x{A}^\text{ref,s}),\delta_{\text{amp}}(\xnf^\text{i},\x{A}^\text{i})) }
\label{eq:d}
\end{equation}
where $p_{\text{amp}}^A$ is the average distance between the time series from one cycle to the class $A$ in the training data, using the $\delta_{\text{amp}}$ feature vector and \textit{ref,s} is the s:th reference cycle. 

The length of the pairwise distance feature vector $\delta_\text{pd}$ depends on the number of classes in the training data, and the number of different distance measures used, plus possible additional features such as the impact frequency of the rock drill. Once these are chosen the vectors length is constant, and also significantly shorter than time series the vector. This makes the vector more suitable for training and evaluation using for instance a Support Vector Machine (SVM) classifier. Using the pairwise distance feature vectors $\delta_\text{pd}$ generated from training data, a classifier is trained. For this work SVM is chosen.

\section{Results}
This section covers the classification results for described methods using the relative measures of difference. The performance of 1NN-DTW, SVM and InceptionTime for classifying single cycle data is shown in Table~\ref{table:accuracy}. Each classifier is evaluated using the different features as well as raw time series. For SVM, which is not suitable for high dimensional inputs, corresponding parts of the $\delta_\text{pd}$ vector are used. This enables using single features $\delta_{\text{amp}}$ or $\delta_{\text{ts}}$, but also combinations of several features. Hence SVM, $\delta_{\text{amp}}$ means SVM using the part of $\delta_\text{pd}$ in~(\ref{eq:delta_pd}) originating from $\delta_{\text{amp}}$ etc. Column \textit{Same} shows the results when training and evaluation is performed on the same individual, though different impact cycles. Column \textit{Different} shows the results when training uses data from one individual, and evaluation is done on the other five individuals. In both cases there is an equal number of samples from each class, hence accuracy is a suitable performance metric. 

\begin{figure}
\includegraphics[width=\columnwidth]{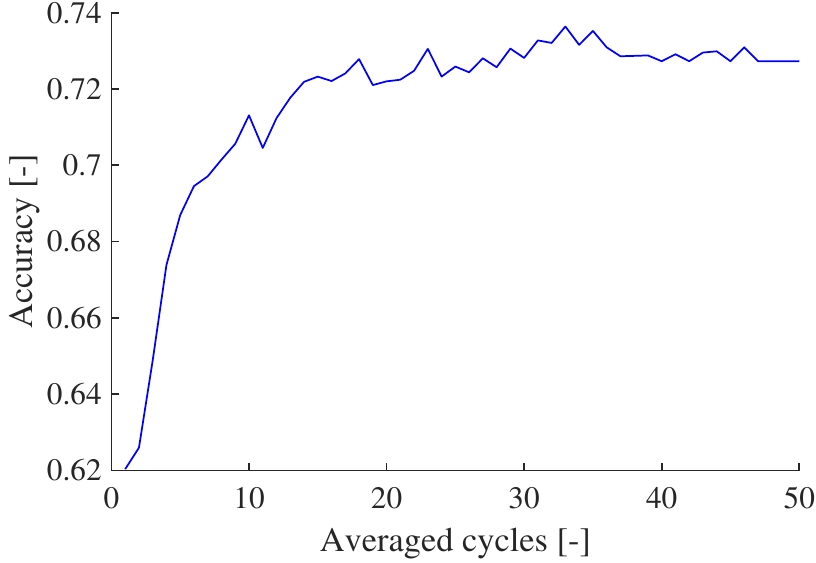}
  \caption{Averaging over consecutive cycles for the InceptionTime, $\delta_{\text{ts}}$. For classifiers with high performance, performance is increased.}
\label{fig:mult_stroke_acc}
\end{figure}

Since impact cycles occur in a time span much shorter than required for faults to develop, additional performance can be gained by averaging the classification over a number of consecutive cycles. Fig.~\ref{fig:mult_stroke_acc} shows how performance increase for the InceptionTime, $\delta_{\text{ts}}$ classifier, as batches of increasing size are used. Table~\ref{table:accuracy_when_averaged} lists the result of the various methods when such averaging is performed over many impact cycles from a particular class, i.e. were the curve in Fig.~\ref{fig:mult_stroke_acc} has leveled off.  

Time series measures are clearly most effective on the same individual, but suffer when used of different individuals. Relative features are less efficient on the same data, but see a much smaller loss of accuracy when evaluated on different individuals. Best overall performance is achieved using Inception Time in combination with the $\delta_{\text{ts}}$ feature.

\begin{table}
\caption{Accuracy for the methods when evaluated on the same or different individuals as during training. Note how out-of-the-box methods using the raw time series excel when training and test data is from the same individual. As individual changes are introduced, relative features outperform time series. The top result is achieved using a complex classifier, Inception time, together with the $\delta_{\text{ts}}$ feature.}
\centering
\begin{tabular}{|c||c|c|}
\hline
Measured quantity 	& Same 	& Different \\
\hline\hline
1NN-DTW, time series								& 92\% 		& 35\% \\
\hline
SVM, time series										& 93\%		& 35\% \\
\hline
InceptionTime, time series								& 99\% 		& 34\% \\

\hline
1NN-DTW, $\delta_{\text{amp}}$						& 73\%		& 40\% \\
\hline
SVM, $\delta_{\text{amp}}$							& 56\%		& 40\% \\
\hline
InceptionTime,  $\delta_{\text{amp}}$					& 79\% 		& 53\% \\

\hline
1NN-DTW,  $\delta_{\text{ts}}$						& 71\%		& 52\% \\
\hline
SVM, $\delta_{\text{ts}}$								& 26\%		& 27\% \\
\hline
InceptionTime, $\delta_{\text{ts}}$						& 89\% 		& 62\% \\

\hline
SVM, $\delta_{\text{amp}}$ and $\delta_{\text{ts}}$		& 63\%		& 43\% \\
\hline
SVM, $\delta_{\text{amp}}$ and $\delta_{\text{ts}}$ + f 	& 65\%		& 50\% \\
\hline
\end{tabular}

\label{table:accuracy}
\end{table}

\begin{table}
\caption{Accuracy for the methods as the average of multiple consecutive cycles is calculated. The relative ordering among methods are not affected.}
\centering
\begin{tabular}{|c||c|c|}
\hline
Measured quantity 	& Same 	& Different \\
\hline\hline
1NN-DTW , time series								& 100\% 		& 35\% \\
\hline
SVM, time series										& 100\%		& 31\% \\
\hline
InceptionTime, time series								& 100\% 		& 33\% \\
\hline

1NN-DTW, $\delta_{\text{amp}}$						& 100\%		& 49\% \\
\hline
SVM, $\delta_{\text{amp}}$							& 82\%			& 49\% \\
\hline
InceptionTime,  $\delta_{\text{amp}}$					& 91\% 			& 72\% \\
\hline

1NN-DTW,  $\delta_{\text{ts}}$						& 82\%			& 65\% \\
\hline
SVM, $\delta_{\text{ts}}$								& 29\%			& 27\% \\
\hline
InceptionTime, $\delta_{\text{ts}}$						& 100\% 		& 73\% \\
\hline

SVM, $\delta_{\text{amp}}$ and $\delta_{\text{ts}}$		& 91\%			& 49\% \\
\hline
SVM, $\delta_{\text{amp}}$ and $\delta_{\text{ts}}$ + f 	& 91\%			& 58\% \\
\hline

\end{tabular}

\label{table:accuracy_when_averaged}
\end{table}

\begin{figure*}
  \includegraphics[width=\textwidth]{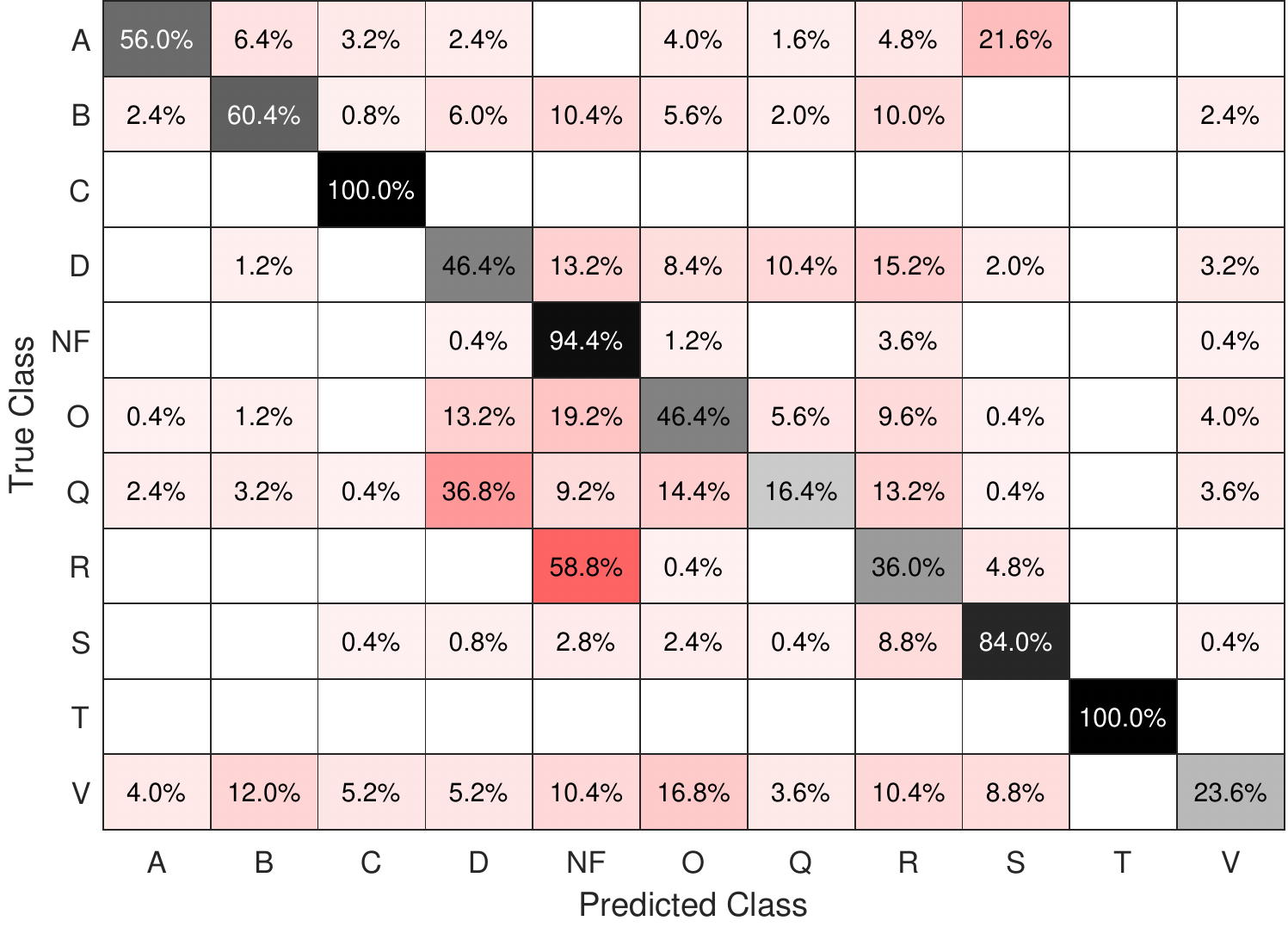}
  \caption{Confusion matrix for the Inception time classifier using $\delta_{\text{ts}}$ features, evaluated on singles impact cycles. Faults Q and D show some mix up, which can be reasonable since both affect the damper circuit and could manifest similarly in the percussion pressure. Fault R is to a large extent confused with NF, which can be interpreted as that this fault does not have much visible effect on the system. In general, most of the faults can be distinguished despite quite small differences in pressure signatures.}
\label{fig:inception_CM}
\end{figure*}

A confusion matrix for the best performing classifier-feature combination on different individuals is presented in Fig.~\ref{fig:inception_CM}. This example was trained using one individual, and evaluated using the other five. The confusion matrix shows the correct answer to be in majority for nine of the eleven cases. The relation between the correct class and the mis-classification give information on how difficult the specific fault is to detect. For C, S, and T, classification is efficient, suggesting these faults are easier to detect. Inspection of the time series confirms these three show a larger difference from NF compared to other faults. The NF case is also clear, which is positive to avoid false alarms. Since the NF case is also measured for each individual, NF could be evaluated separately.  The expected performance for NF vs. not NF would then be similar to the \textit{Same} column in Table~\ref{table:accuracy_when_averaged}.    

\begin{figure}[t]
\centering
\begin{subfigure}[t]{\columnwidth}
\includegraphics[width=\columnwidth]{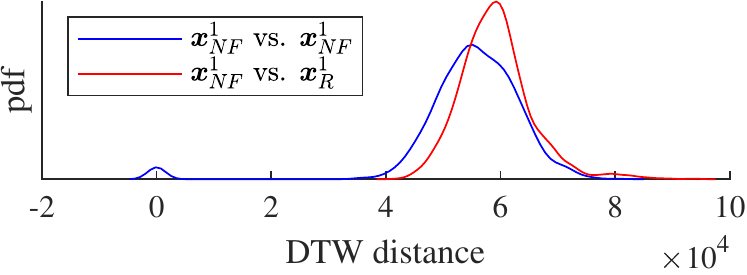}
\caption{Distribution of DTW distances between time series of classes R and NF shows too much overlap to be separated by a threshold.}
\label{fig:caseR_a}
\end{subfigure}
\begin{subfigure}[t]{\columnwidth}
\includegraphics[width=\columnwidth]{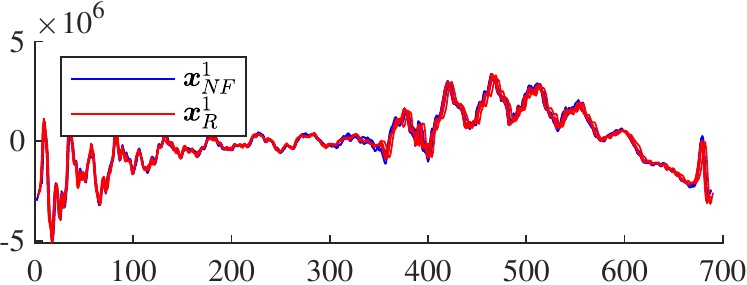}
\caption{The time series show very little difference for fault class R.}
\label{fig:caseR_b}
\end{subfigure}
\caption{The two classes NF and R are almost identical, showing the difficulty of this classification task.}
\end{figure}

Fault R, the broken return accumulator is interesting. From Fig.~\ref{fig:inception_CM}, a majority of cycles from fault R are mis-classified as NF. If NF could be ruled out using alternative methods, the R case would obtain a clear majority of the remaining samples. An attempt to do so using DTW on the time series directly and assuming NF data is available from each individual is shown in Fig.~\ref{fig:caseR_a}. The idea is that a clear gap should exist between the two distributions for this indivual, and hence a threshold on DTW distance to the NF-class could distinguish R from NF for any individual. Fig.~\ref{fig:caseR_a} shows no such gap, and the conclusion is that classes NF and R simply are to similar to be classified correctly with this method. This is emphasized in Fig.~\ref{fig:caseR_b}, where the time series from the same individual are close to identical and thus very difficult to classify correctly. For faults Q and V, which are spread out over many classes, the weak majority for V, or even minority for Q indicate that these faults are also difficult to classify correctly.   


Experiments were also performed to assure no great difference occur if another of the individuals are chosen as reference. For the InceptionTime, $\delta_{\text{ts}}$ such alteration affect the accuracy between 0.52-0.62, were the lowest performance is found for individuals most different from the rest. This indicates there is still improvements to be made on finding the ideal $\delta$ that is not affected by individual differences. 	 

\section{Discussion / Implementation}
The concept of removing individual bias showed to be very efficient for scalar hand crafted features, and is a natural way to handle individual variations. But how to do such compensation for time series data, and especially when there is large impact from wave propagation, is not a trivial extension. A common notion in applied machine learning is that with enough data insights can be delivered without much need for domain expertise. The performance when using state-of-the-art classifiers on time series data within the same individual also give this impression. But as more variations are introduced, domain knowledge play an increasingly important role in selecting algorithms, generating features and pre-processing of data. This is an important property to keep in mind 

The features $\delta_{\text{ts}}$ and $\delta_{\text{amp}}$ showed to be closer than raw time series to the idealized feature $\delta$ described in (\ref{eq:rel_features_1}) and (\ref{eq:rel_features_2}). An interesting extension to this idea of using reference data, and possibly an improvement, would be to learn this feature $\delta$ directly from data. A possible starting point could be Siamese networks \citep{chicco2021siamese}, \citep{bromley1993signature}, that specialize in learning a notion of difference from a reference example, typically utilized for face recognition and similar applications. With such setup, the need to generalize over all different individuals would be reduced, and instead the model would learn how different faults manifest in relation to the reference data. The reference could then be exchanged for any other individual on deployment. 

\begin{figure}[t]
\centering
\includegraphics[width=\columnwidth]{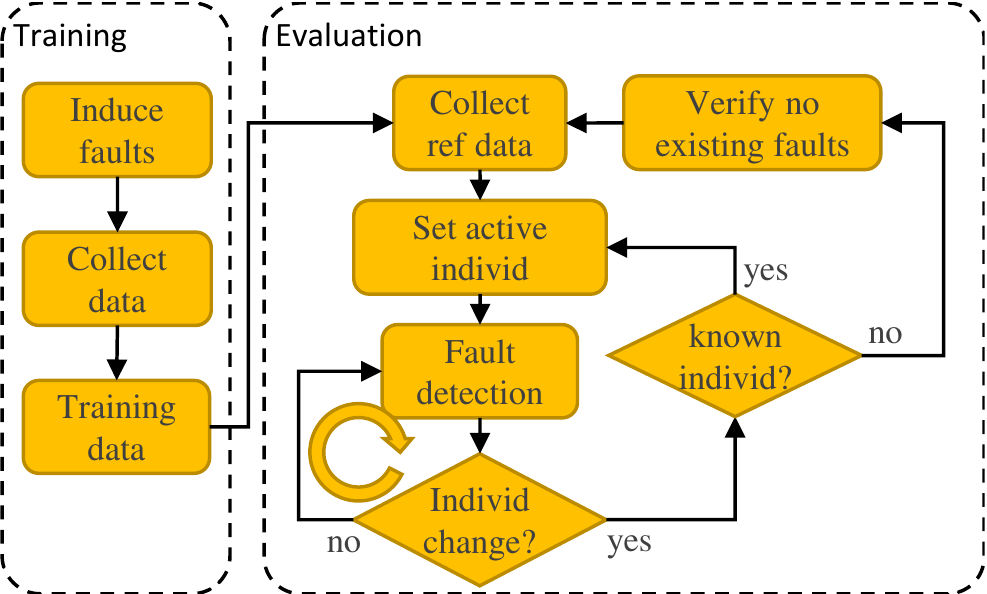}
\caption{The proposed process for training and evaluation. Training takes place off line. On deployment, a No-fault reference measurement is performed on the current individual, to be used for calibration. During usage of the equipment, only \textit{Fault detection} occurs, together with means for detecting a change of configuration.}  
\label{fig:process}
\end{figure}

The improvement from using reference data do come at a cost. The need for individual reference data makes implementation more involved, compared to a system where all generalization over individuals is covered by offline training data. Fig.~\ref{fig:process} shows a principal way of handling individual difference upon deployment. For a static system, where the individual does not change over time, the implementation is straightforward and reference data is collected only once upon commissioning. Potential problems occur if there is a change, which would need to first be detected, and then remedied by activating a new set of reference data. If the new individual is unknown, new reference data would need to be collected. The really difficult part here is that none of the faults should be present when doing so. 
 
\section{Conclusions}
The work is summarized by reconnecting to the original research questions.

RQ1: How can classification be used when the target system is different from the training system? 
In the broader sense, if there is no similarity, classification is not possible. What is shown in this paper is a method to make the unseen target system more similar to the training data by using modified features. This can reduce some of the need to collect data spanning all possible variations, a task that is often a major hurdle for fault identification tasks.  

RQ2: Can No-fault data from the target system be used to reduce the effect of individual differences? 
For scalar features, a single bias compensation showed to be sufficient to adapt a hand crafted feature. For wave propagation data, a difference measure based on DTW showed considerable improvements. The key is to find the type of relative information actually affected by the different faults. 

RQ3: For this class of problems, how well do state-of-the-art Time Series Classification methods work?
For a case where the evaluation data can have significant differences, the tested out-of-the-box methods struggles. Having data from a large section of available variations is typically a prerequisite for data driven machine learning models, and when this conditions is not met the methods naturally fails. 

RQ4: What role does domain expertise have to make a machine learning scheme useful? 
Using state-of-the-art methods on available data, even though the data is selected by domain experts to capture as much information as possible, does not necessarily give good results. It is shown that by adding domain knowledge, such as the benefits of relative features targeting amplitude and time shift specifically, can have a large impact on the performance even when using the same classifier. 

By using relative features, significant improvements can be seen in classification results on this problem. The benefits are most prominent in systems where inducing faults for many individuals is difficult, but collecting nominal data from the deployed system is easy. Assuming this is a common property in fault detection applications allows for many interesting automatic relative feature extraction methods in the future.

\section*{Acknowledgements}
The authors would like to thank Peder Haraldsson and Martin Persson, Epiroc Rock Drills AB, for assistance during data collection in this work. This work was partially supported by the Wallenberg AI, Autonomous Systems and Software Program (WASP) funded by the Knut and Alice Wallenberg foundation.

\bibliography{mybib}             

\end{document}